\documentclass[reprint, english,aps,prl,bibnotes]{revtex4-1}
\usepackage{graphicx}
\usepackage[T1]{fontenc}
\usepackage[latin9]{inputenc}
\usepackage{amstext}
\usepackage{amssymb}
\usepackage{multirow}
\usepackage{dcolumn}
\usepackage{color}
\usepackage{amsmath}
\usepackage{babel}
\makeatletter

\begin{document}

\title{Strong environmental coupling in a Josephson parametric amplifier}

\author{J.Y. Mutus}
\thanks{These authors contributed equally to this work}
\affiliation{Department of Physics, University of California, Santa Barbara, California
93106-9530, USA}
\author{T.C. White}
\thanks{These authors contributed equally to this work}
\affiliation{Department of Physics, University of California, Santa Barbara, California 93106-9530, USA}
\author{R. Barends}
\affiliation{Department of Physics, University of California, Santa Barbara, California 93106-9530, USA}
\author{Yu Chen}
\affiliation{Department of Physics, University of California, Santa Barbara, California 93106-9530, USA}
\author{Z. Chen}
\affiliation{Department of Physics, University of California, Santa Barbara, California 93106-9530, USA}
\author{B. Chiaro}
\affiliation{Department of Physics, University of California, Santa Barbara, California 93106-9530, USA}
\author{A. Dunsworth}
\affiliation{Department of Physics, University of California, Santa Barbara, California 93106-9530, USA}
\author{E. Jeffrey}
\affiliation{Department of Physics, University of California, Santa Barbara, California 93106-9530, USA}
\author{J. Kelly}
\affiliation{Department of Physics, University of California, Santa Barbara, California 93106-9530, USA}
\author{A. Megrant}
\affiliation{Department of Physics, University of California, Santa Barbara, California 93106-9530, USA}
\author{C. Neill}
\affiliation{Department of Physics, University of California, Santa Barbara, California 93106-9530, USA}
\author{P.J.J. O'Malley}
\affiliation{Department of Physics, University of California, Santa Barbara, California 93106-9530, USA}
\author{P. Roushan}
\affiliation{Department of Physics, University of California, Santa Barbara, California 93106-9530, USA}
\author{D. Sank}
\affiliation{Department of Physics, University of California, Santa Barbara, California 93106-9530, USA}
\author{A. Vainsencher}
\affiliation{Department of Physics, University of California, Santa Barbara, California 93106-9530, USA}
\author{J. Wenner}
\affiliation{Department of Physics, University of California, Santa Barbara, California 93106-9530, USA}

\author{K.M. Sundqvist}
\affiliation{Electrical and Computer Engineering, Texas A\&M University, College Station, Texas 77843, USA}

\author{A.N. Cleland}
\affiliation{Department of Physics, University of California, Santa Barbara, California 93106-9530, USA}
\author{John M. Martinis}
\email{martinis@physics.ucsb.edu}
\affiliation{Department of Physics, University of California, Santa Barbara, California 93106-9530, USA}

\date{\today}
\begin{abstract}

We present a lumped-element Josephson parametric amplifier designed to operate with strong coupling to the environment.  In this regime, we observe broadband frequency dependent amplification with multi-peaked gain profiles. We account for this behaviour using the ``pumpistor'' model which allows for frequency dependent variation of the external impedance. Using this understanding, we demonstrate control over gain profiles through changes in the environment impedance at a given frequency. With strong coupling to a suitable external impedance we observe a significant increase in dynamic range, and large amplification bandwidth up to 700 MHz giving near quantum-limited performance.  

\end{abstract}
\maketitle

Parametric amplification is a result of frequency mixing via a nonlinear element coupled to an external environment.  Amplifiers based on this principle  have achieved near quantum limited performance\cite{caves:qNoise}, essential for high fidelity measurement of both optical \cite{kumar:opticalql} and microwave \cite{axion, wallraff:dispersive} signals.  In the microwave domain the Josephson parametric amplifier \cite{yurke:originalJPA,yamamoto:fluxDrive,castellanos:JPA,hatridge:LJPA,abdo:jpc,ourPaper} (JPA) has enabled new studies of quantum jumps \cite{vijay:jumps} and measurement of quantum trajectories \cite{hatridge:backaction}.  While well suited to single qubit dynamics, progress in scaling to larger quantum algorithms and  fault-tolerant quantum computing \cite{fowler:surface, barends:gates, day:broadband} is limited by JPA bandwidth and dynamic range.  JPA performance is constrained by weak coupling between the nonlinear resonator and the environment, chosen to simplify amplifier dynamics.

In this Letter, we experimentally demonstrate a JPA operated in a previously unreported regime of strong coupling to the environment. Physical insight into this coupling interaction follows naturally from the ``pumpistor'' model of a flux driven SQUID, previously used to describe only degenerate (phase-sensitive) amplification \cite{sundqvist:pumpistor}.  By adapting the non-degenerate (phase-preserving) ``pumpistor'' theory \cite{sundqvist:pumpistor2} we create a model which accounts for dramatic improvements in both quantum limited bandwidth and dynamic range observed in this device. Additionally, by varying the environment we demonstrate significant control over amplifier dynamics and provide a further verification of the full ``pumpistor'' theory.

The JPA relies on the Josephson inductance to create a nonlinear resonator which is typically weakly coupled to a 50\,$\Omega$ embedding environment.  When driven by a pump tone of sufficient power, energy is coupled from the pump ($\omega_p$) into other signals within the resonator bandwidth.  A signal applied near the resonant frequency ($\omega_o$) results in an amplified signal ($\omega_s$) and idler ($\omega_i$) tone. In this work, the amplifier is operated as a non-degenerate (phase-preserving) three-wave mixing amplifier, where $\omega_p = \omega_s + \omega_i$ and $\omega_p \approx 2 \omega_o$. The amplifier operates in a reflection mode where a microwave circulator separates the incoming signal from the outgoing amplified signal and idler tones, which are further amplified by a cryogenic following amplifier, typically a  high-electron mobility transistor (HEMT) amplifier \cite{bradley:HEMT}. 

\begin{figure}
  
  \centering
    \includegraphics[width=0.45\textwidth]{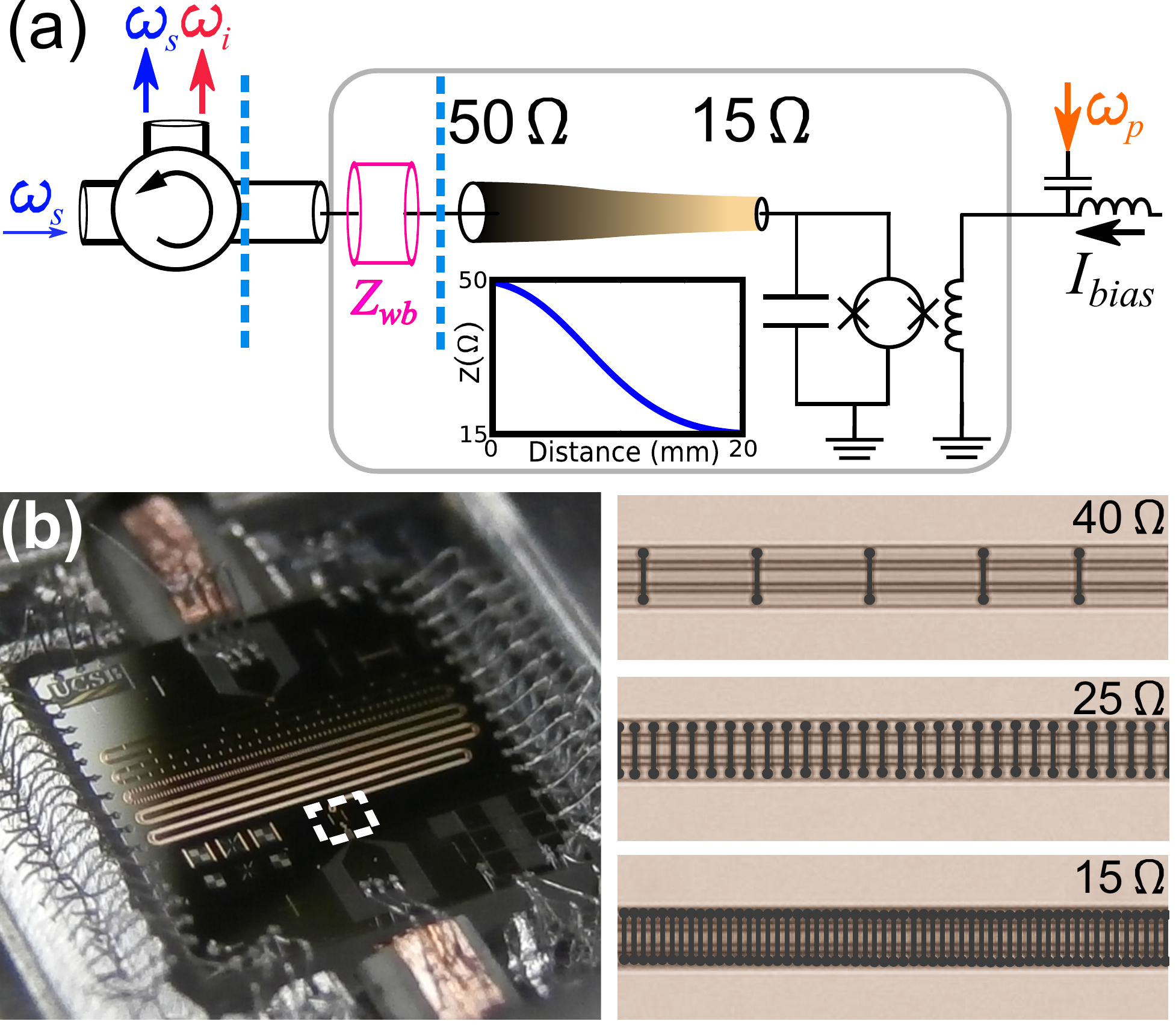}
  
\caption{(Color Online) Schematic and photograph of the IMPA. \textbf{(a)} Circuit diagram of the IMPA (gray, light box) and accessory microwave hardware.  The device consists of a nonlinear LC resonant circuit coupled to the 50\,$\Omega$ environment by a tapered transmission line, smoothly varying from 50\,$\Omega$ to 15\,$\Omega$, with profile shown in the inset. The bias port (right) injects current $I_\text{bias}$ to change the resonant frequency $\omega_o$ of the circuit by varying the inductance of the SQUID loop.  Dominant sources of reflections (dashed blue lines) in the chain are the circulator, used to separate incoming and amplified outgoing signals, and the wire-bond (pink, light) shown as $Z_\text{wb}$.  \textbf{(b)} Photograph of device.  The chip is a 3\,mmX3\,mm square. The gradient of the crossover density is visible as the 20\,mm long CPW becomes increasingly bronze (light) in color. The LC resonant circuit is contained within the dashed box. On the right are photo-micrographs of different regions of the hybrid CPW/microstrip transmission line. As the density of 2\,$\mu m$ wide crossovers (false color in black) increases, the impedance of the transmission line drops.}
\label{fig:device}
\end{figure}

The impedance-transformed parametric amplifier (IMPA) shown in Fig.\,\ref{fig:device} builds on a typical lumped-element style JPA consisting of a SQUID loop with 100\,pH of combined geometric and non-linear (Josephson) inductance $L_j$  shunted by a 4\,pF parallel plate capacitor $C$, for a characteristic impedance of $1/\omega_oC \approx 5\,\Omega$. Typically, this resonator is coupled to a 50\,$\Omega$ transmission line, either directly or with a coupling capacitor. Practical measurements with the JPA are limited by narrow bandwidth (10-20\,MHz typ.) and the low signal power (-120\,dBm typ.) at which the amplifier saturates.  These figures of merit are many orders of magnitude lower than the HEMT following amplifier. Bandwidth in the JPA scales as $1/Q$ and saturation power scales as $I_{c}^2/Q^3$ \cite{manucharyan:microEmbedding} where $I_c$ is the critical current of the SQUID and $Q = Z_o\omega_o C = Z_o/\omega_o L$ is the low power coupled $Q$ of the JPA.  For a fixed environment impedance $Z_o$ and frequency $\omega_o$, the coupled $Q$ and critical current cannot be varied independently because $L_j \propto 1/I_c$, introducing trade-offs between saturation power and bandwidth. This trade off can be circumvented using multiple SQUIDS in series \cite{eichler:paramps}, but this makes fabrication less reliable and complicates device operation.

In the IMPA, we instead transform the environmental impedance $Z_o$, increasing coupling, lowering $Q$ and thus simultaneously increasing the bandwidth and saturation power. We use a tapered impedance transformer (Fig.\,\ref{fig:device}) to lower the effective external impedance seen by the JPA from 50\,$\Omega$ to about 15\,$\Omega$. In this way, we can directly probe the effects of lowering the coupled $Q$ while at the same time increasing both saturation power and available bandwidth.

A hybrid geometry was adopted for the taper, since the 15\,$\Omega$ to 50\,$\Omega$ impedance range is not intrinsically suited to either a purely co-planar waveguide (CPW) or microstrip transmission line. The tapered line consists of a fixed geometry CPW shunted with parallel plate capacitor cross-overs. The sections with a cross-over approximate a microstrip transmission line, with much lower local characteristic impedance. The small size of the crossovers (2\,$\mu$m) relative to the wavelength of a 6\,GHz photon, allows us to vary the impedance smoothly with the density of crossovers, following a 20\,mm long Klopfenstein taper \cite{klopfenstein:taper,pozar}, a profile chosen to minimize the pass-band ripple of the network; see Fig.\,\ref{fig:device} \cite{supp}.

\begin{figure}
  
  \centering
    \includegraphics[width=0.45\textwidth]{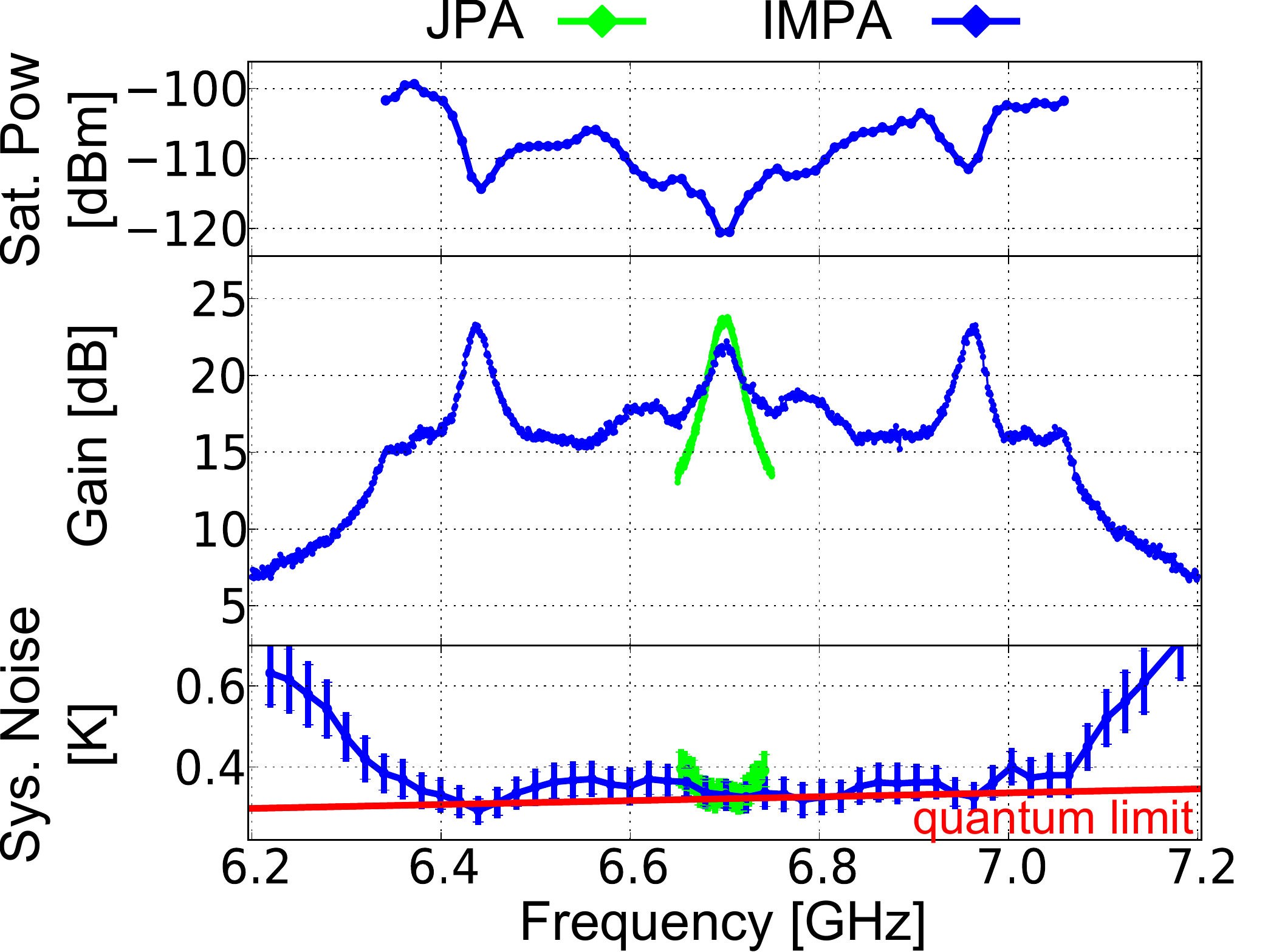}
\caption{(Color Online) Input saturation power, gain, and noise performance of the IMPA. Here we show amplifier performance centered at 6.7\,GHz. The IMPA (blue, dark) provides an average input saturation power (defined as 1 dB compression point) of $-108\,\textrm{dBm}$ with regions as high as $-103$\,dBm.  The device gives greater than 15\,dB of gain and  near quantum-limited performance over a bandwidth of nearly 700 MHz. This is compared to a typical JPA with a $Q=10$ (green, light) for a similar gain, which only provides quantum-limited performance over a 60 MHz band.  We define the quantum limit as one photon $\hbar\omega$ of total system noise at the input of the amplifier.  Here deviations from the quantum noise limit correspond to gains that are unable to completely overwhelm the noise added by the HEMT following amplifier.  Error bars on the noise correspond to potential systemic calibration errors on HEMT system noise.}
  \label{fig:performance}
\end{figure}

Using this new device we measure a significant increase in saturation power, the power at which the gain compresses by 1\,dB, with values as high as $-103$\,dBm at 15\,dB gain, as shown in Fig.\,\ref{fig:performance}. Decreasing the coupled $Q$ has the added benefit of increasing bandwidth. We have measured amplification bandwidths of nearly 700\,MHz, shown in Fig.\,\ref{fig:performance} for data centered about 6.7 GHz. Due to the multi-peaked gain features visible in the figure, we define the amplification bandwidth as the frequency range over which the device approaches the quantum noise limit.  We calculate system noise using the method of signal to noise ratio improvement \cite{hatridge:LJPA,ourPaper} over the calibrated noise of the HEMT following amplifier.

In both cases the measured improvement differs from theoretical predictions. The increase in saturation power, while significant, is about a factor of 3 less than the factor $\sim 30$ expected from the change in coupled $Q$. This could be a result of the increased bandwidth, corresponding to an increase of amplified quantum noise, which consumes a portion of the higher saturation power. The factor of 10 improvement in the bandwidth also cannot be explained by the reduced coupled $Q$. Moreover, the shape of the gain profile differs significantly from the typical Lorentzian described by most resonant JPA models \cite{siddiqi:JBA, manucharyan:microEmbedding, eichler:paramps} and thus requires a detailed understanding of how the JPA interacts with variations in the microwave environment.

For a JPA coupled to a perfect 50\,$\Omega$ environment, the incident signal tone at $\omega_s$ reflects off the LC resonator, where pump photons are converted to amplified signal $\omega_s$ and idler $\omega_i$ tones. In the case of an imperfect match to the environment, the outgoing amplified signal and idler tones are back-reflected towards the JPA, creating standing waves and affecting device gain through constructive or destructive interference. These standing waves are properly thought of as variations in the frequency dependent admittance $Y_\text{ext}(\omega_s)$ (inverse impedance) of the environment seen by the JPA. The predominant sources of reflections in the microwave chain are due to the wire-bond and microwave circulator interfaces, shown as dashed lines in Fig.\,\ref{fig:device} (a). Typically, experiments are designed to minimize the distance between the JPA and these reflection planes, spacing out these standing waves in frequency. As a result, the effects of these standing waves are not so apparent in devices with $Q$>10 as variations in the impedance are small over the response bandwidth of the JPA.

The full ``pumpistor'' theory \cite{sundqvist:pumpistor2}, while previously validated for only the degenerate frequency (phase-sensitive) case \cite{sundqvist:pumpistor}, is well suited to a detailed analysis of the effect of these standing waves on non-degenerate gain in a JPA.  Here, the non-linearity of a flux-pumped SQUID loop is treated as a power dependent modification of the SQUID inductance.  For a signal at frequency $\omega_s$ the admittance of the loop becomes $Y(\omega_{s}) = 1/i \omega_{s} L_{0} + 1/i \omega_{s} (L_{1}+L_{2})$, where the three elements of the inductance are
\begin{eqnarray}
L_\text{0} &=& L_{j}/\cos(\pi \Phi_{Q}/\Phi_{0}) \label{eq:L0}\\
L_\text{1} &=& -\dfrac{4 L_j \cos(\pi \Phi_{Q}/\Phi_{0})}{\pi^{2} \sin^{2}(\pi \Phi_{Q}/\Phi_{0})} \left( \dfrac{\Phi_{0}}{\Phi_{ac}}\right)^{2} \label{eq:L1}\\
L_\text{2} &=&  i\dfrac{4 \omega_{i} L_j^{2} Y_\text{ext}^\star(\omega_i)}{\pi^{2} \sin^{2}(\pi \Phi_{Q}/\Phi_{0})} \left(\dfrac{\Phi_{0}}{\Phi_{ac}}\right)^{2}. \label{eq:L2}
\end{eqnarray}
Here the idler frequency $\omega_i = \omega_p-\omega_s$, $L_{j} = \Phi_0/(2 \pi I_c)$ is the unbiased SQUID inductance, $\Phi_{Q}$ is the DC flux bias, and $\Phi_{ac}$ is the amplitude of the flux pump.  The dependence of $L_2$ on the external admittance at the idler frequency $Y_\text{ext}^\star(\omega_i)$ comes about because the pump also drives oscillations at the idler frequency, and the magnitude of these oscillations depends on the output admittance.  As the pump power increases from zero, $L_{1}+L_{2}$ emerges as an element in parallel with the initial SQUID inductance $L_0$. The term $L_{1}$ modifies the inductance of the circuit, lowering the operating frequency as pump power increases. The term $L_{2}$ represents an imaginary inductance that gives rise to a negative real impedance given by $ \text{Re}[i \omega_{s} L_{2}]$.

As the JPA is a reflection amplifier, we can use our ``pumpistor'' model to calculate the reflection coefficient and thus the gain $G$ using the admittance (impedance) mismatch between the external evironment $Y_\text{ext}$ and the paramp admittance $Y_\text{JPA}$. 
\begin{equation}
\label{eq:gain}
G(\omega_s) = \dfrac{Y_\text{ext}(\omega_s)-Y_\text{JPA}(\omega_s)}{Y_\text{ext}(\omega_s)+Y_\text{JPA}(\omega_s)}.
\end{equation}
Using Eqs. (\ref{eq:L0})-(\ref{eq:L2}) we derive a simplified approximation of the JPA admittance which includes the SQUID loop and shunt capacitance $C$
\begin{equation}
\label{eq:y_jpa}
Y_\text{JPA}(\omega_s) = \dfrac{-2i(\omega_o-\omega_s)}{\omega_o\omega_s L_c} - \dfrac{\pi^2 \sin^2(\pi \Phi_Q/\Phi_0)(\Phi_{ac}/\Phi_0)^2}{ 4 \alpha  \omega_s \omega_i L_\text{j}^2 Y_\text{ext}^\star(\omega_i)},
\end{equation}
where $1/L_c = 1/L_0 + 1/L_1$ is the combined parallel inductance of the SQUID, $\omega_o = 1/\sqrt{CL_c} \approx \omega_p/2$, and $\alpha = 1+Q^2 \approx 10$ is due to a series to parallel circuit conversion \cite{supp}.  Equations (\ref{eq:gain}) and (\ref{eq:y_jpa}) describe how a knowledge of the frequency dependent admittance of the environment at both the signal and idler frequencies is required to model amplifier behavior. 


\begin{figure*}[t]

  \centering
    \includegraphics[width=\textwidth]{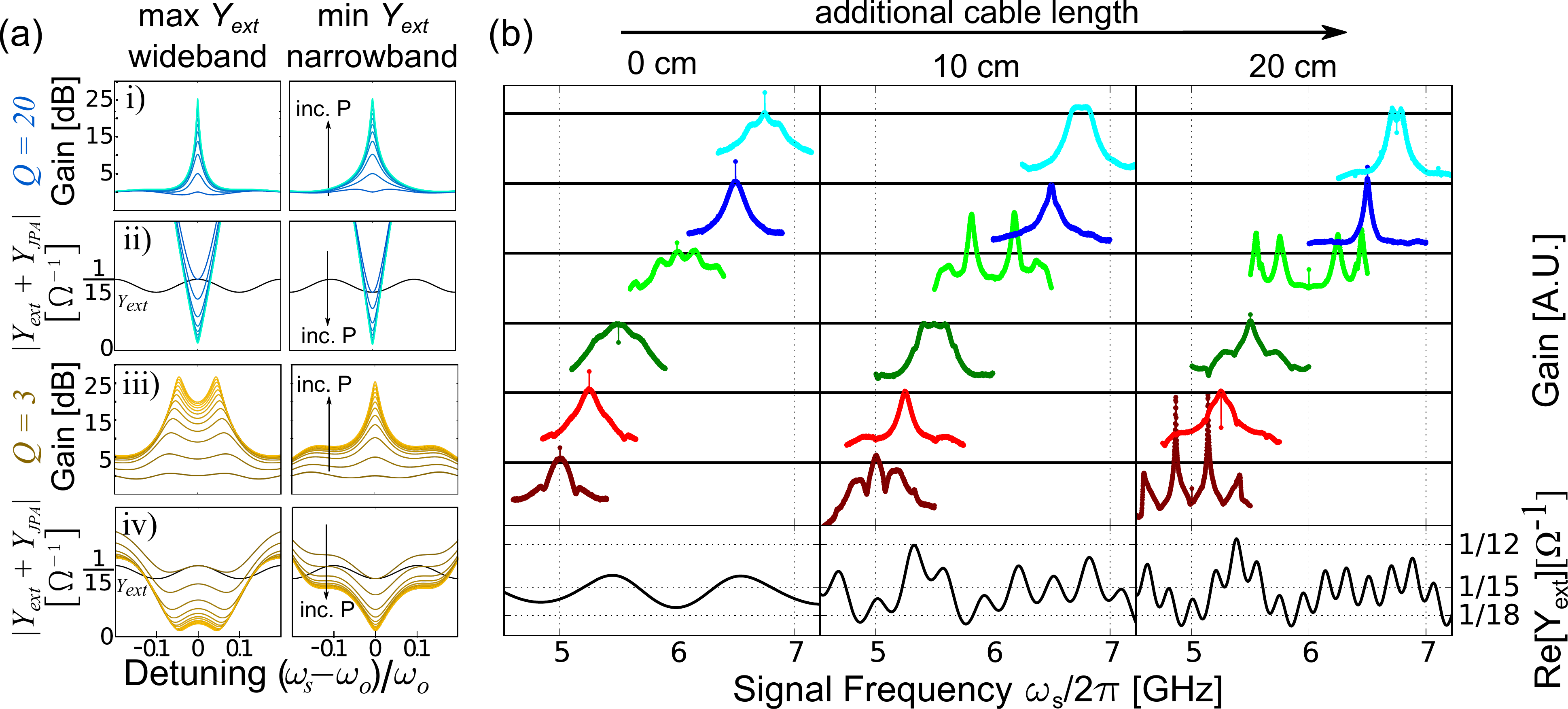}
 
  \caption{(Color Online) Effect of variation in external admittance on amplifier performance. \textbf{(a)} Simulated gain profiles (i,iii) showing the effect of external impedance on gain for both weak ($Q=20$) and strong ($Q=3$) coupling. Higher pump power is denoted by lighter colored traces and indicated by the arrow. The denominator of Eq.\,\ref{eq:gain} (ii,iv) is plotted, with the gain shown in (i,iii). As pump power increases the denominator is displaced from $Y_\text{ext}$ towards zero by an increasingly negative $Y_\text{JPA}$. In a weakly coupled device (i,ii) the gain profile is not affected by the shape of the external admittance. In the case of strong coupling (iii, iv) the response of the JPA is broader and the gain profile is greatly influenced by the external admittance. If $\omega_p/2 = \omega_o$ is centered at a maxima in external admittance the gain is broadened (left), if $\omega_o$ is centered on a minima it is narrowed (right).  \textbf{(b)} Experimental variation of the standing waves in the external environment. Gain profiles of approximately 20\,dB, and offset by 20\,dB (denoted by bold horizontal lines) are shown as the amplifier frequency $\omega_o$ is tuned from 5 to 7\,GHz. Standing waves are introduced by lengthening the cable between the IMPA and the circulator; the longer the cable, the more closely spaced in frequency are the standing waves.  Amplitude variations in $Y_\text{ext}$ come from a superposition of standing waves in the cable and on-chip taper. The gain profiles become increasingly featured as the variations in $Y_\text{ext}$ (calculated using SPICE) increase with cable length.}
  
 \label{fig:peaks}
\end{figure*}

We show in Fig.\,\ref{fig:peaks}(a) solutions to Eq. (\ref{eq:gain}) and (\ref{eq:y_jpa}) with a 10\% sinusoidal variation in $Y_\text{ext}$. With increasing power, the denominator in Eqs. (\ref{eq:gain}) is brought closer to zero, resulting in increased gain.  The $Q$ = 20 case shows that the effect of $Y_\text{ext}$ on the overall gain is dominated by the narrowed response of the JPA. For comparison, in the $Q$ = 3 case (gold, light), the response of the JPA is broad enough to sample variations in the external environment. Moreover, the profile of the gain reflects the shape of  $Y_\text{ext}$ at that frequency.  This results in the variations in bandwidth seen in the left versus right simulation. We note that the measured gain profiles show a high degree of symmetry for $\omega_s$ about $\omega_p/2$, regardless of the detailed behavior of $Y_\text{ext}$. This is to be expected, as the gain samples $Y_\text{ext}$ at both the signal and idler frequencies, which are symmetrically placed about $\omega_p/2$. A broader sampling of frequency dependent gain profiles calculated using Eqs. (\ref{eq:gain}) and (\ref{eq:y_jpa}) are shown in Ref.\,\cite{supp} with a similar degree of symmetry.

To test the dependence of amplifier performance on environment admittance, we changed the pattern of standing waves on the output line by changing the length of cable separating the device from the circulator. We measured the gains as a function of $\omega_s$ for different resonant frequencies $\omega_o$ from 5 to 7\,GHz with direct connection to the circulator, a 10\,cm and 20\,cm cable.  The results of this experiment are shown in Fig.\,\ref{fig:peaks}(b).  In each case, a series of gain profiles with $\sim 20$\,dB peak gain are shown, spaced vertically. We also plot the frequency-dependent output impedance taken from simulations using the corresponding length of cable.  This frequency-dependent admittance can be estimated using lumped circuit models to approximate the dominant contributions from the circulator and wirebond connections.

When the amplifier is connected directly to the circulator, the admittance variation is minimized and the IMPA more consistently approaches the expected Lorentzian gain profiles.  When connected using the 10 cm cable a drastic change is exhibited in many of the peaks showing both broadening at some frequencies and narrowing at others.  When the 20 cm cable is used the output impedance varies more rapidly and the device performance becomes increasingly erratic while often exhibiting multiple distinct resonant peaks. The experimental data shows good qualitative agreement to that predicted by the ``pumpistor'' reflection model.  Using the understanding gained from this model, the IMPA has been tuned up for broadband, high-power, multi-qubit readout in existing experiments \cite{jeffrey:readout,barends:gates} with a roughly 15\,cm copper cable at the operation point shown in Fig.\,\ref{fig:performance}.

In conclusion, we have demonstrated validation of the ``pumpistor'' theory and application of a new model for understanding parametric amplifier behavior.  In the strongly coupled limit this model predicts unexpectedly large bandwidths, which have been observed in the IMPA with near quantum-limited noise performance.  Using this model, further improvements should be possible by shaping the external embedding impedance, possibly with alternate matching networks \cite{ranzani:matching}.  This large bandwidth, along with a significant increase in saturation power, has allowed us to study high power measurement in a multi-qubit device \cite{jeffrey:readout}, suitable for error correction architectures \cite{fowler:surface, barends:gates}.

This work was supported by the Office of the Director of National Intel-ligence (ODNI), Intelligence Advanced Research Projects Activity (IARPA), through the Army Research Office grant W911NF-10-1-0334. All statements of fact, opinion or conclusions contained herein are those of the authors and should not be con-strued as representing the official views or policies of IARPA, the ODNI, or the U.S. Government.  Devices were made at the UC Santa Barbara Nanofabrication Facility, a part of the NSF-funded National Nanotechnology Infrastructure Network, and at the NanoStructures
Cleanroom Facility. 

\bibliographystyle{apsrev}
\bibliography{references}

\end{document}


\title{Supplementary Information}

\date{\today}

\section{Supplementary Information}

\subsection{LJPA Fabrication}

\begin{figure}[b]
  \centering
    \includegraphics[width=0.7\textwidth]{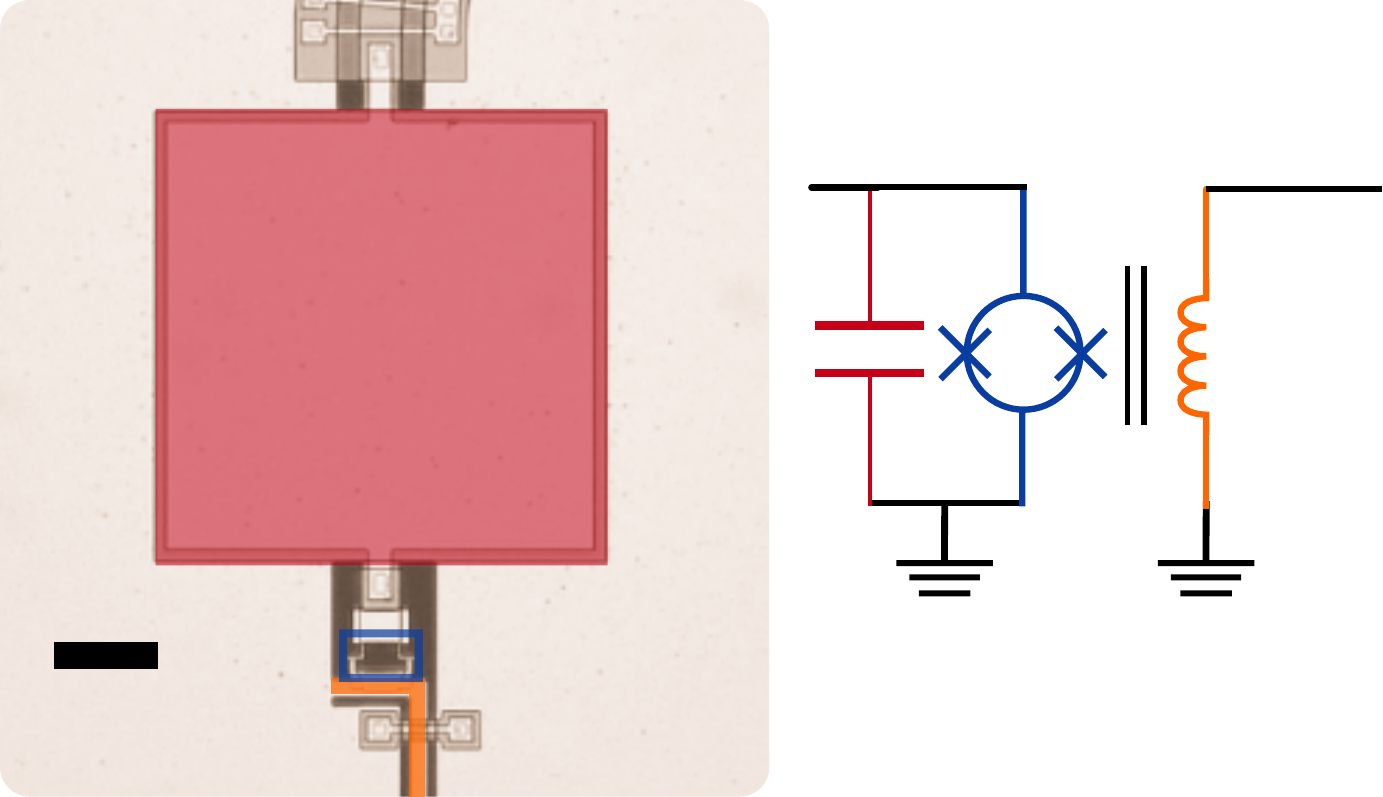}
  \caption{Optical micro-graph and circuit diagram of the lumped-element JPA design used in the IMPA.  The device consists of a 4\,pF parallel plate capacitor (red), in parallel with a 4\,$\mu$A ciritical current SQUID loop (blue box). The pump tone and bias current are applied via an on-chip bias line (orange). The mutual inductance between the SQUID and bias line is 2.6\,pH.  Scalebar is 25\,$\mu$m.}
  \label{fig:circuit}
\end{figure}

The resonant circuit of the IMPA is based on a previous experiment with lumped-element JPAs coupled directly to a 50 ohm environment \cite{ourPaper}. This design is shown in Fig. \ref{fig:circuit}, along with a circuit diagram for the device.  The circuit consists of 5 separate layers of optical lithography as well as a final e-beam lithography step to deposit the junctions.  First, a base layer of 100 nm of aluminum was sputtered onto a sapphire substrate and patterned optically to form the bulk of the circuit.  The insulator (250 nm amorphous silicon) was then deposited and patterned to create vias between the top wiring and base wiring layers.  After an in-situ argon ion mill to remove the native oxide from the base layer to ensure good electrical contact between base and top wiring, a 100 nm of aluminum top-wiring was sputter deposited.  The top wiring and insulator layers were then patterned and etched to form the parallel plate capacitor of the resonator and the crossovers.  Finally e-beam lithography was used to pattern the junctions, which were deposited using double angle evaporation and liftoff  in an e-beam evaporator.

\subsection{Taper Design}

\begin{figure}
  \centering
    \includegraphics[width=0.5\textwidth]{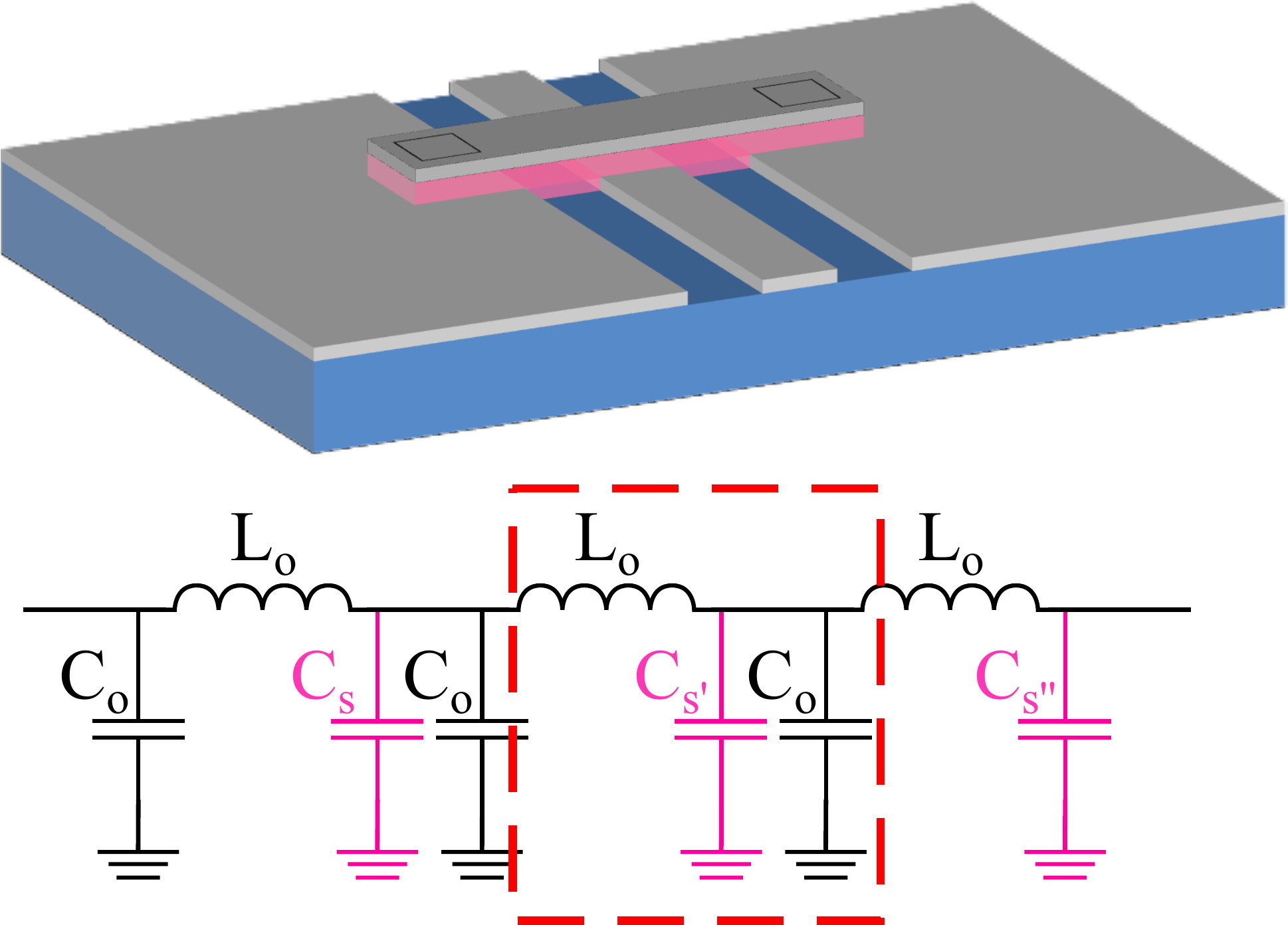}
  \caption{The hybrid CPW/microstrip geometry. To match of a wide range of impedances with features defined by photolithography, the CPW with 10-5\,$\mu$m center trace-gap widths was shunted by a variable density of shunt capactitor crossovers.  As the crossover width is much lower than a wavelength, the capacitance per unit-length of the transmission line smoothly increases, continuously reducing the characteristic impedance from 50\,$\Omega$ to 15\,$\Omega$. A unit cell for this taper in the dashed red box, is shown above.}
  \label{fig:taper}
\end{figure}

We chose a Klopfenstein taper \cite{klopfenstein:taper,pozar} to transform the impedance from 50\,$\Omega$ to 15\,$\Omega$ over a large bandwidth while minimizing the pass-band ripple. Given the 1\,$\mu$m limitations of our optical lithography process, a taper consisting entirely of a 15\,$\Omega$ co-planar wave-guide (CPW) transmission line would require a 200\,$\mu$m center-trace with a 1\,$\mu$m gap width. For a microstrip geometry we were contrained to a low impedance imposed by our thin a-Si dielectric, which would require a 100\,nm trace width for a 50\,$\Omega$ line.  A hybrid CPW/microstrip transmission line shown in Fig. \ref{fig:taper} was adopted for the taper where a 10\,$\mu$m center trace and 5\,$\mu$m gap CPW was shunted by a variable density of 2\,$\mu$m wide microstrip sections.  The result is a 20\,mm long tapered transmission line designed for a maximum reflection of -20\,dB above 4 GHz.

The microstrip part of this geometry was created using multi-layer crossovers as additional shunt capacitance.  We tested this model using microwave finite-element simulations to ensure that the shunt capacitance behaved as expected, and to correct for the change in inductance imposed by the crossovers.  After extensive simulation we found that using a small crossover width of 2\,$\mu m$ and varying the density of crossovers provided the best control over impedance, as it allowed more evenly distributed shunt capacitance, prevented the inductance correction from becoming too large, and allowed the microstrip sections be applied uniformly over the meandered CPW.

\subsection{Time Domain Reflectometry at milliKelvin temperatures}

\begin{figure}
  \centering
    \includegraphics[width=0.9\textwidth]{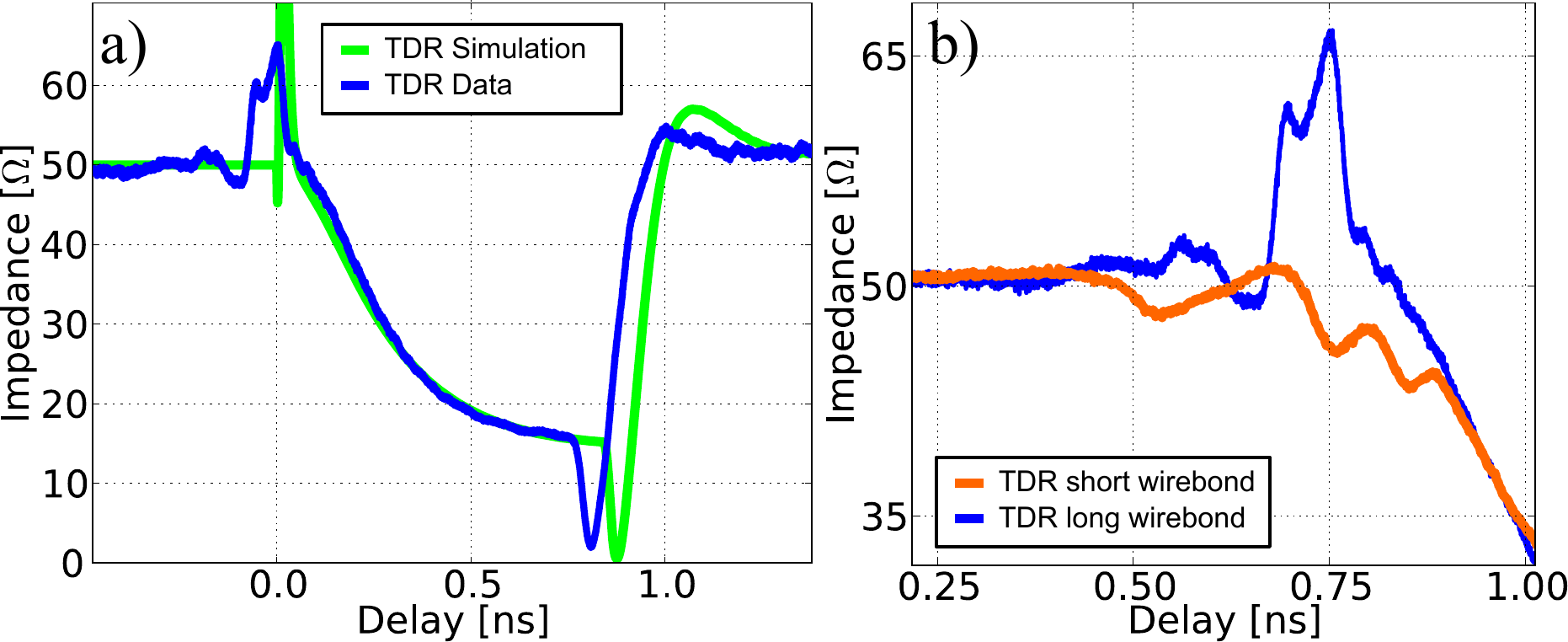}
  \caption{TDR analysis of an IMPA sample.  (a) The TDR response of experiment (at 50 mK) \textit{vs.} theory (SPICE simulations) of the IMPA.  The tapered sections agree well.  Both curves end in a dip followed by a jump in impedance, corresponding to the lumped element resonator.  The peaks seen before the taper begins correspond to an impedance mismatch at the wire bond due to a large series inductance $\sim$ 1\,nH.  This mismatch can lead to standing waves on the taper which limit IMPA performance.  (b) An expanded view of the wire-bond mismatch for a sample with long (1-1.5\,mm) vs short (0.3-0.5\,mm) wire bonds.  The improved wire bond mismatch is similar in magnitude to that of an SMA connector seen at 0.5\,ns.}
  \label{fig:tdr}
\end{figure}

To verify that the profile of the taper followed the designed value, we employed time domain reflectometry (TDR) \cite{cole:TDR}. For this measurement, the time dependent reflection of a system is measured after the application of a fast step pulse, in order to measure the impedance of the system as a function of delay time, equivalent to the distance along the line.  These measurements, shown in Fig. \ref{fig:tdr}, were carried out in an adiabatic demagnetization refrigerator (ADR) with a base temperature of $\sim$55\,mK to ensure the aluminum was superconducting, otherwise the loss from the normal aluminum CPW would overwhelm the response TDR.  All TDR data was taken using a Tektronix DSA8300 Digital Serial Analyzer with a 80E08B TDR/Sampling Module.  The sampling module was connected directly to a CuNi line down to 4K followed by a direct NbTi line down to the IMPA at 50 mK. Since the excitation voltage of this particular TDR sampling head is fixed, at 10\,dB attenuator was applied to reduce the current at the device. The attenuated TDR data were calibrated using a 50\,$\Omega$ terminator and an open at room temperature and corrected to obtain the proper impedances.

These TDR measurements were used to minimize reflections at the wire-bond due to excess inductance.  As shown in Fig. \ref{fig:tdr} excess inductance in the wire-bonds of the signal line can lead to a large impedance mismatch at the start of the taper.  This mismatch can lead to standing waves which severely limit the usable frequency range of the IMPA.  It was only after greatly reducing the length of the wire-bonds from 1-1.5\,mm to 0.3-0.5\,mm that we observed the enhanced performance reported here.

\subsection{Pumpistor Model}

\begin{figure*}
  \centering
    \includegraphics[width=.9\textwidth]{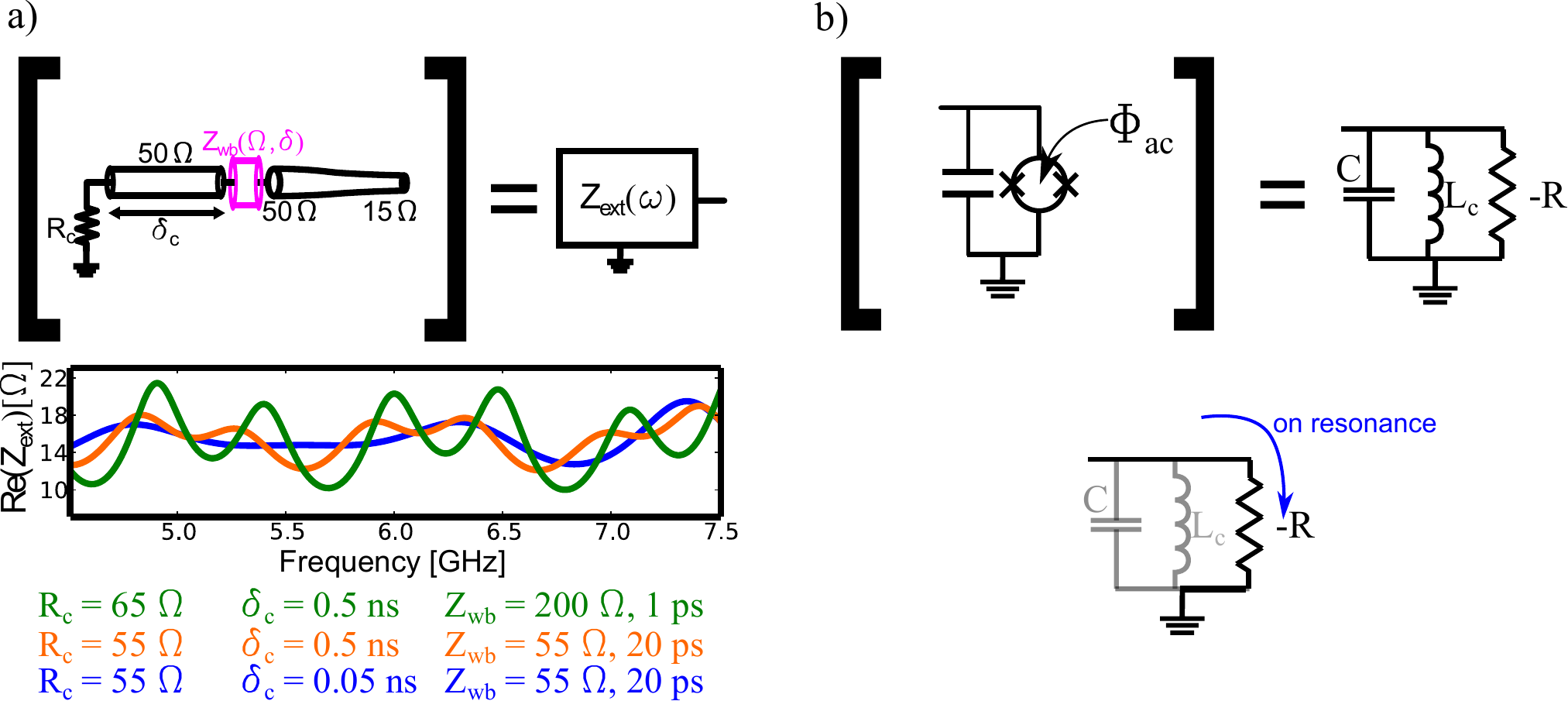}

  \caption{Simulated frequency dependent performance for the Pumpistor model. a) The impedance of the external circuit was calculated using SPICE. The taper, wirebond and distance between the IMPA and circulator were modeled using transmission line segments of variable impedance and delay. The magnitude of the reflection due to the circulator was modeled as a resistor to groud with a mis-match corresponding to the VSWR specification for the circulator. b) The JPA can be modeled as a linear circuit element using the pumpistor model. This circuit can be approxmated as a parallel LC resonator shunted by a negative resistance. On resonance, the current through the JPA will be shunted through the negative resistance and the impedance will be given by the total real component $-R$. }
  \label{fig:yjpa}
\end{figure*}

To understand the non-Lorentzian gain peaks and enhanced bandwidth in the IMPA, one must properly model the interaction of the JPA with the frequency dependent impedance environment. Since this is a reflection amplifier, the gain can be calculated by the reflection coefficient at the interface between the external circuitry and the JPA. The frequency dependent impedance of the external circuit can be modeled straightforwardly using a SPICE model with the parameters shown in Fig. \ref{fig:yjpa}a. The tapered transmission line was modeled using 40 sections of equal delay transmission line corresponding to the impedance profile calculated using Ref. \cite{pozar}. The mismatch due to the wire-bond was obtained using a section of transmission line at the input of the taper with a variable impedance and delay. The effect of the reflections due to the circulator was modeled using a resistor to ground whose value corresponds to the range of voltage standing wave ratio (VSWR) given by the manufacturer (Quinstar) specifications, at the end of a 50 $\Omega$ transmission line with variable delay. 

\begin{figure*}
  \centering
    \includegraphics[width=.5\textwidth]{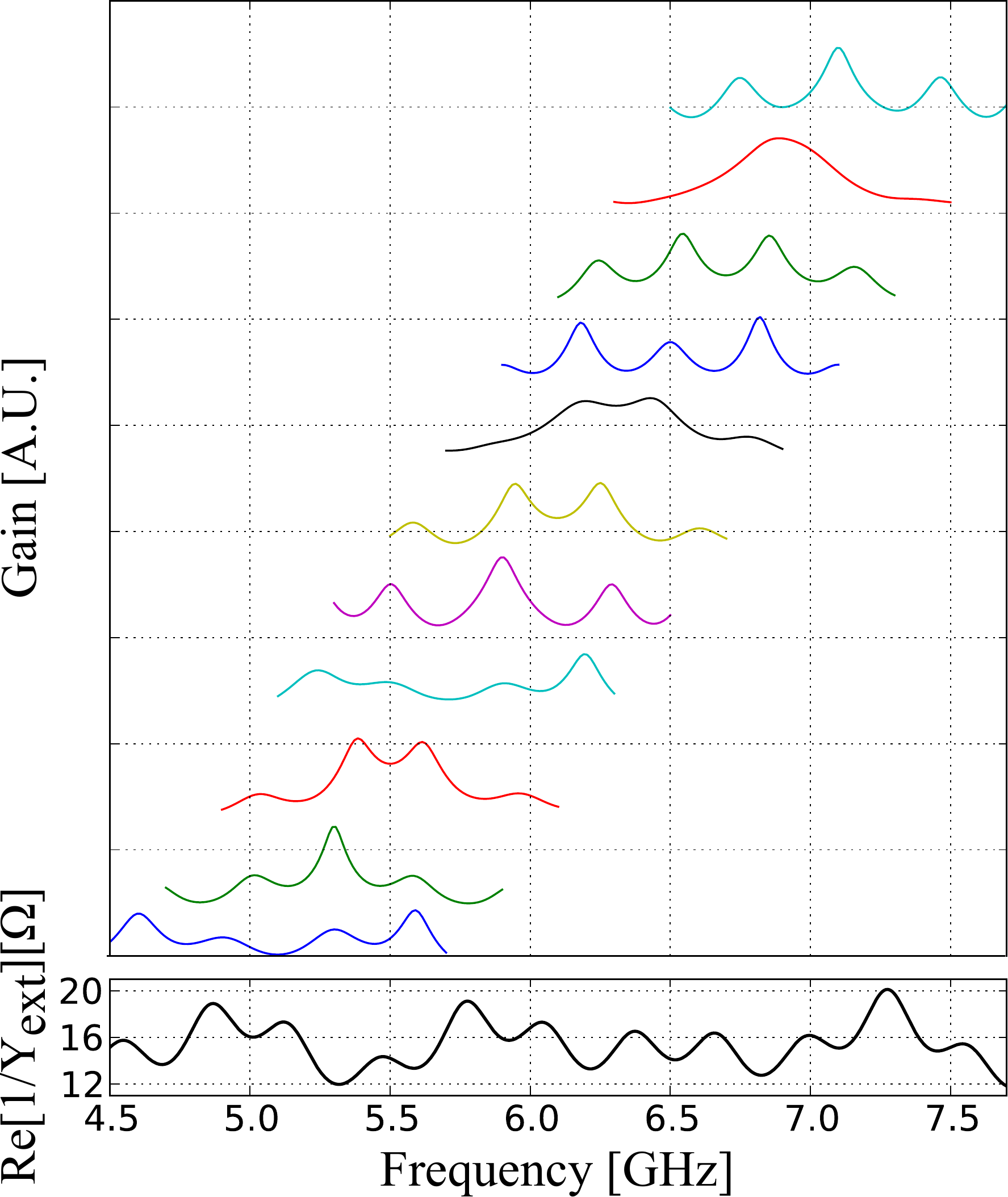}

  \caption{Additional simulations of the pumpistor model. Frequency dependence of the calculated gain profiles (each with a max gain approximately 20\,dB)  for $Y_\text{ext}$ corresponding to the 10\,cm cable length shown in  Fig. 3(b) of the main paper. These calculated gain profiles are mostly symmetric, with regimes of single, multiple and broadened peaks, and show good qualitative agreement with the data. The vertical gridlines are spaced by 20\,dB. }
  \label{fig:sims}
\end{figure*}

\begin{figure*}
  \centering
    \includegraphics[width=1.\textwidth]{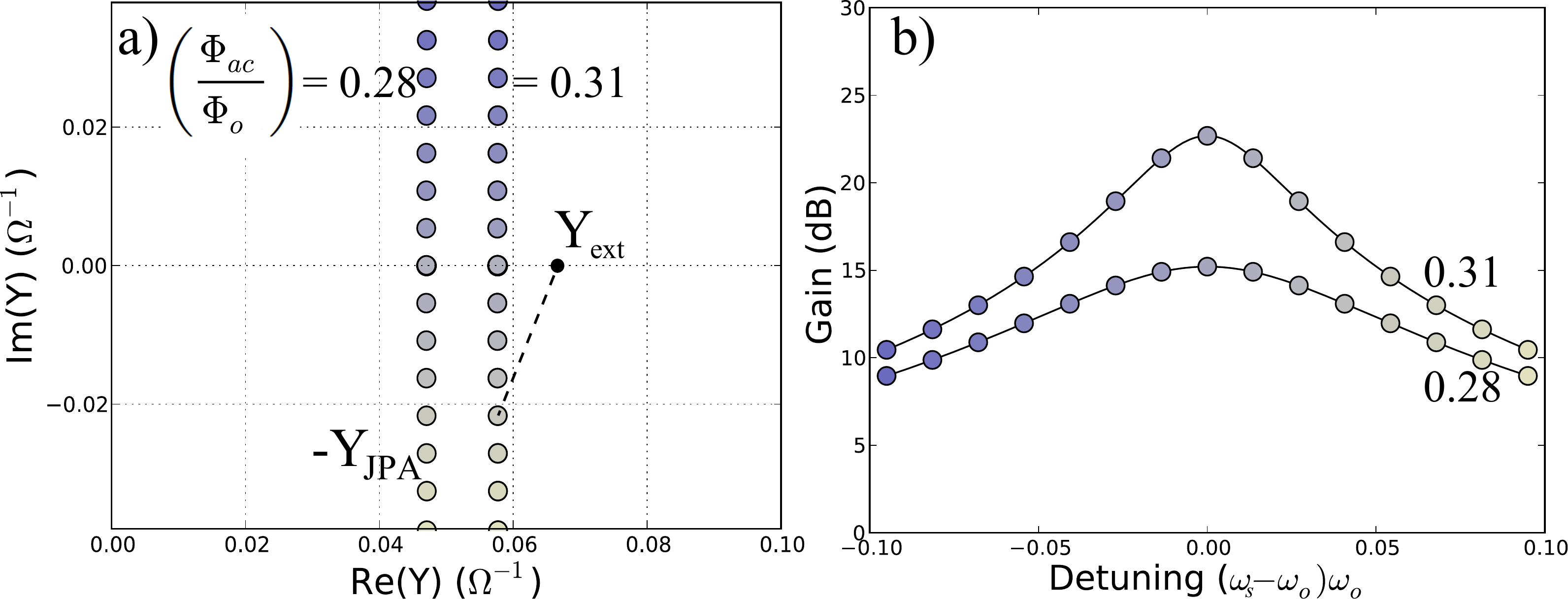}

  \caption{The frequency dependent behavior of $-Y_\text{JPA}$ (open, colored circles) and $Y_\text{ext}$ (black dot) in the complex plane.  (a) For a constant $Y_\text{ext}$, $Y_\text{JPA}$ is shown for two pump powers over a frequency range from  $\omega_o-0.1\omega_o$ to $\omega_o+0.1\omega_o$ (each colored dot corresponds to a gain given in in panel (b)) where $\omega_o=\omega_p/2$. (b) The gain is inversely proportional to the distance between $Y_\text{JPA}$ and $Y_\text{ext}$ for a given frequency, indicated by the dashed line in (a). Since $Y_\text{ext}$ is constant with frequency, the gain varies with the distance between the line given by $Y_\text{JPA}$ and a point defined by $Y_\text{ext}$. The highest gain is found on resonance where  $Y_\text{JPA}$ is closest to  $Y_\text{ext}$.  The typical Lorenzian gain profiles are shown in (b) corresponding to the two different powers: $\Phi_{ac} / \Phi_o$ = 0.28 and 0.31. Higher pump powers push $-Y_\text{JPA}$ to the right; as it nears $Y_\text{ext}$ the gain increases and bandwidth decreases.}
  \label{fig:complex}
\end{figure*}

\begin{figure*}
  \centering
    \includegraphics[width=1.\textwidth]{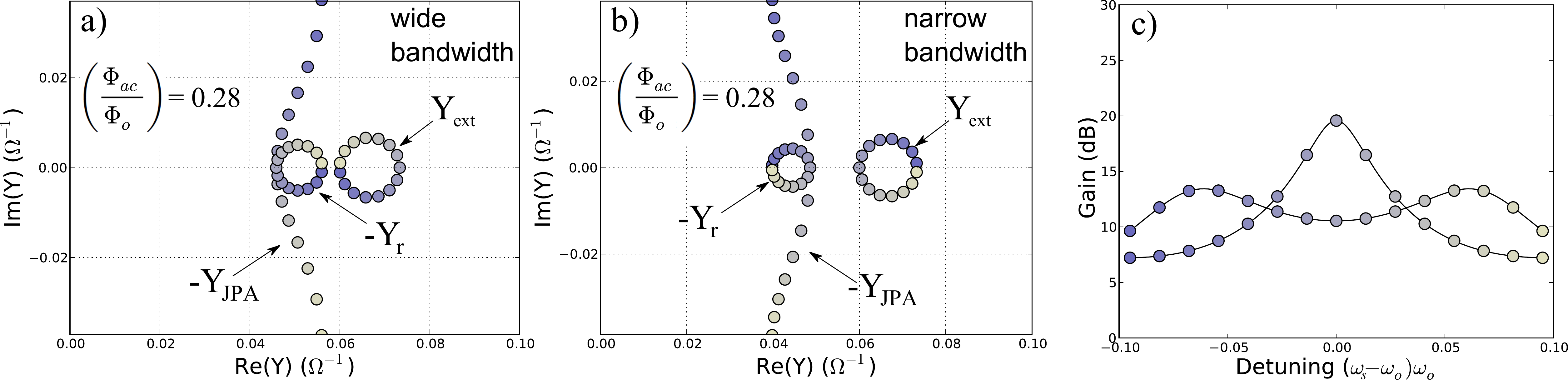}

  \caption{The frequency dependent behavior of $-Y_\text{JPA},-Y_R$ and $Y_\text{ext}$ in the complex plane for a sinunsoidally varying $Y_\text{ext}$.  The frequency at a given point, denoted by its color, each dot corresponds to a gain point shown in panel (c), where $\omega_o=\omega_p/2$. The gain at a given frequency is inversely propotional to the distance between $-Y_\text{JPA}$ and  $Y_\text{ext}$.  (a) In the wide bandwidth case, the profile $Y_\text{JPA}$ nears  $Y_\text{ext}$ over a wider frequency range than otherwise. The resistive term reflects the variation in $Y_\text{ext}$, and displaces $Y_\text{JPA}$, providing a closer match to $Y_\text{ext}$ over a broader range of frequencies than in Fig. \ref{fig:complex}. In the narrow bandwidth case  (b) $Y_\text{JPA}$ approaches $Y_\text{ext}$ over a  smaller range of freuqnencies.  This happens when $\omega_p/2$ is located at a local minimum of $Y_\text{ext}$. The rolloff in gain with detuning is due to the imaginary portion of Eq. (\ref{eq:final}) growing with increased detuning. The corresponding gain profiles for (a) and (b) are shown in (c).}
  \label{fig:complex-pi}
\end{figure*}

The pumpistor model allows the flux-pumped SQUID to be treated as a linear cicuit component with the three inductances, given by Eqs. (1), (2), and (3) in the main text, where $L_0$ is the unbiased SQUID inductance in parallel with $L_1$ and $L_2$ which modify this bare inductance as pump power increases.  To simplify the circuit analysis we express the effects of $L_1$ and $L_2$ as an equivalent parallel circuit with  $L_1'$ and $L_2'$, such that $L_1'$ modifies only the inductance of the circuit. Gain is introduce by the imaginary inductance of $L_2'$ which behaves like a negative resistance, coupling power from the pump into the circuit.  The parallel equivalent circuits are given by
\begin{eqnarray}
Q_s  &=& \dfrac{\omega_s L_1}{- i \omega_s L_2} = \dfrac{1}{\omega_i[L_j /\cos(\pi \phi_q / \phi_0)]Y_\text{ext}^*(\omega_i)},\\
L_1' &=& L_1 \left( 1 + 1/Q_s^2\right),\\
L_2' &=& L_2 \left(1+Q_s^2\right).
\end{eqnarray}
For resonant circuits here we use $Q_s= Q \approx 3$, so $L_1' \sim L_1$ and $L_2' = \alpha L_2 \approx 10 L_2$ to keep track of the series to parallel conversion. This allows us to write down an approximate circuit for the JPA shown in Fig. \ref{fig:yjpa}, where $C$ is the shunt capacitance of the parallel plate capacitor of the LJPA, $L_c$ is an effective inductance including $L_0$ and $L_1$ for a given bias point and $\alpha L_2$ gives a negative resistance in parallel. The circuit for these parallel admittances can be written as
\begin{equation}
Y_\text{JPA} = i \omega_s C + \dfrac{1}{i \omega L_c} - Y_R,\label{eq:admins}
\end{equation}
where $Y_R$ is the magnitude of the JPA response given by:
\begin{equation}
Y_R = \dfrac{\pi^2 \sin^2(\pi \Phi_Q/\Phi_0)}{4 \alpha \omega_s\omega_iL_\text{j}^2 Y_\text{ext}^*(\omega_i)} \left(\dfrac{\Phi_{ac}}{\Phi_{0}}\right)^{2}.\label{eq:res}
\end{equation}
Simplifying Eqs. (\ref{eq:admins}) with $\omega_o^2 = 1/L_cC$:
\begin{eqnarray}
Y_\text{JPA}&=&\dfrac{1}{i\omega_sL_c}(1-\dfrac{\omega_s^2}{\omega_o^2})-Y_R \\
&=&\dfrac{1}{i\omega_sL_c}\dfrac{(\omega_o-\omega_s)(\omega_o+\omega_s)}{\omega_o^2}- Y_R .\label{eq:final}
\end{eqnarray}
With $\omega_o + \omega_s \approx 2\omega_o$, this yields Eqs. (5) from the main text. 

This parallel to series conversion used to obtain Eq.\,(\ref{eq:final}) implicitly assumes $Q_s$ is real.  However, when a complex $Y_\text{ext}^*$ is used to compute $Q_s$, the resulting correction to $Y_\text{JPA}$ is negligible. It turns out the largest assumption in obtaining Eq.\,(\ref{eq:final})  is that $Q_s \approx 3$, a value frequency in frequency. For this case $\alpha$ is predicted imprecisely over the narrow range of frequencies near resonance, as plotted in Fig. 3 of the main paper. However, $\alpha$ is simply a  scaling factor for the pump power and $Y_\text{ext}^*$, both of which are not known precisely from experiment.  These expressions are combined with the results of the SPICE model to obtain frequency dependent gain performance with good qualitative agreement to experiment as shown in Fig \ref{fig:sims}. Only with careful and very precise measurement of  reflections in the microwave chain can one yield  quantitative agreement with the exact formulation. 

For additional insight into the engineering of large bandwidth, interaction between the terms  $Y_\text{ext}$,  $Y_\text{JPA}$ and $-Y_R$  (calculated using Eq.  (\ref{eq:final}))  in the complex plane is shown in Figs. \ref{fig:complex} and \ref{fig:complex-pi}. The gain is inversely proportional to $|Y_\text{ext} + Y_\text{JPA}|$, equivalent to the distance between $Y_\text{ext}$ and  $Y_\text{JPA}$. By increasing pump power, $-Y_R$ increasingly deforms $Y_\text{JPA}$. Large bandwidth is achieved by minimizing the distance between $Y_\text{JPA}$  and $Y_\text{ext}$ over a larger range of frequencies.  This is shown in contrast to the ideal case with no variation in $Y_\text{ext}$, which produces the typical Lorentzian gain profile.

\newpage

\bibliographystyle{apsrev}
\bibliography{references}